# DEVELOPMENT OF AN OPEN EDUCATION RESOURCES (OER) SYSTEM: A COMPARATIVE ANALYSIS AND IMPLEMENTATION APPROACH

## Nimol Thuon[1,2*], Wangrui Zhang[2]

*[1]Department of Information and Communication Engineering, Institute of Technology of Cambodia, Cambodia*
*[2]Faculty of Management, Economics and Social Sciences, University of Cologne, Germany*

*Corresponding Author*

## ABSTRACT

*Several institutions are collaborating on the development of a new web-based Open Education Resources (OER) system designed exclusively for non-commercial educational purposes. This initiative is underpinned by meticulous research aimed at constructing an OER system that optimizes user experiences across diverse user profiles. A significant emphasis is placed on utilizing open-source tools, frameworks, and technologies. The project includes a comparative analysis of the top five open-source Learning Management Systems (LMS), providing critical insights to inform the development process. The primary objective is to create a web-based system that facilitates the sharing of educational resources for non-commercial users, leveraging information and communication technologies. The project is structured around two key teams: a research team and a development team. This comprehensive approach is intended to establish a robust, user-centric OER system, informed by insights from existing platforms and the latest advancements in open education resource development.*

## 1. INTRODUCTION

In response to the growing demand for accessible educational resources, many educational institutions are experiencing an increase in student enrollment. This surge has heightened the need for educational materials. In today's rapidly evolving society, the relationship between knowledge, education, and learning has been fundamentally reshaped by the pervasive influence of Information and Communication Technologies (ICT). Consequently, the demand for accessible and adaptable educational materials has led to the emergence of Open Educational Resources (OER) as a transformative solution.

The UNESCO Institute for Information Technologies in Education (IITE) has been actively supporting UNESCO Member States in policy-making and national capacity building, focusing on effective ICT integration within educational systems and teaching processes. The institute's mission is to bridge the digital knowledge gap by enabling access to information, fostering scientific research, sharing educational practices, and facilitating self-education.

At its core, OER encompasses a broad spectrum of educational resources that are openly available for use by educators and students, eliminating the need for royalties or license fees. These resources include learning content, software tools for content development and distribution, and implementation resources, such as open licenses. Global advocacy for OER has gained significant traction, recognizing its pivotal role in achieving educational goals at all levels. The UNESCO World Open Educational Resources Congress in 2012 produced a declaration urging countries to foster awareness and use of OER, develop enabling environments for ICT, reinforce strategies and policies, promote open licensing frameworks, support capacity building, and encourage collaboration and research in OER development.

The term "Open Educational Resources" was formally coined at a UNESCO conference in 2002, defined as the open provision of educational resources enabled by ICT for consultation, use, and adaptation by a non-commercial community of users. This definition has since evolved to encompass a wide array of learning content, tools, and implementation resources. Embracing OER represents a strategic response to the evolving landscape of education, reflecting a paradigm shift towards open, accessible, and adaptable educational materials for the benefit of learners and educators worldwide.

## 2. CHALLENGES

Despite the increasing availability of Open Educational Resources (OERs), their integration into daily teaching practices at higher education institutions remains limited. Conventional textbooks and readings continue to dominate teaching materials, even as the majority of students engage in online learning. Course management systems primarily serve as platforms for sharing syllabi, class notes, general communications, and grade tracking, yet their potential for facilitating the widespread use of OERs remains largely untapped. Educators have expressed a growing need for more openly accessible resources for teaching, learning, and research, as the current availability of open-source documents falls short of their requirements. This shortfall hinders their ability to adapt to diverse learning needs and preferences.

Recognizing the vital importance of OERs and the pressing need for expanded educational resources, numerous institutes and organizations worldwide have embarked on creating and promoting Open Educational Resources. The realization of the significant advantages offered by OER systems has underscored the critical necessity for educators to access a wider range of resources for more efficient and effective learning. As the traditional educational landscape undergoes dynamic changes, the demand for adaptable, accessible, and diverse educational materials has become increasingly urgent. This urgency has prompted a collective shift towards embracing OERs as a fundamental component of modern education.

## 3. OBJECTIVE

Our decision to establish an Open Education Resources (OER) system for educational institutions has set the objective for our project. This initiative aims to enable users to upload documents, manage users, oversee learning materials (including e-learning and document tutorials), and engage in discussions and forums related to learning. During my internship at educational institutions, my responsibilities included researching existing popular OER systems worldwide to identify new functionalities, user types, processing technologies, and conducting a survey to compare Learning Management System (LMS) tools and semantic search engines suitable for implementation in educational settings.

The development of this project requires robust material management, allowing instructors and learners to upload documents or video learning materials. To address this need, we conducted extensive research and surveys to compare the functionalities of various leading LMS tools, carefully selecting those best suited for our project's development. Additionally, recognizing the limitations of general website searches, which typically only allow for queries based on file titles or information, we explored and implemented semantic search functionality to extract top keywords from the text within each file, enhancing the search experience for users.

In developing OER for educational institutions, thorough research into useful tools, frameworks, and development languages is essential. Notably, the Learning Management System (LMS) plays a pivotal role in OER system development. Open-source solutions have emerged as a viable option for numerous reasons. After a comprehensive comparison of five LMS platforms (Moodle, Blackboard, Canvas, Sakai, and Latitude), we determined that these systems offer strong features and usability, making them well-suited for the development of OER systems in educational institutions. Furthermore, the development of a web system necessitates careful consideration of numerous tools, frameworks, and development languages to ensure the creation of a robust and reliable system. We meticulously compared and surveyed top tools and frameworks, considering various essential factors to make informed decisions about their integration into the OER development process in educational institutions.

## 4. COMPARISON OF TOOLS AND FRAMEWORKS
### A. Programming Languages and Frameworks

In our evaluation of programming languages and frameworks for the development of the OER system, we have selected PHP 8 and Laravel 8 for their contemporary features and advantages.

**PHP 8 was chosen due to its numerous benefits, including**
1. **Fast Load Time:** PHP 8 enhances site loading speed, improving user experience and overall performance.
2. **Cost Efficiency:** Most tools associated with PHP are open-source software, reducing licensing costs and promoting accessibility.
3. **Database Flexibility:** PHP is highly flexible for database connectivity, supporting various database systems, with MySQL being the most commonly used.
4. **Increased Available Programming Talent:** PHP's wide adoption results in a larger pool of talent, facilitating modifications and development at a lower cost per hour.

**Laravel 8 was chosen for several reasons, including**
1. **Popularity:** As one of the most popular open-source frameworks for PHP, Laravel 8 enjoys a robust and active community, ensuring ongoing support and development.

2.  **Scalability:** It is capable of handling substantial web development projects, providing flexibility for future expansion and growth.
3.  **Lightweight Template Engine:** Laravel 8 offers a lightweight yet powerful template engine, enabling efficient and dynamic content rendering.
4.  **Modern PHP Principles:** The framework adopts modern PHP principles, promoting cleaner and more maintainable code.
5.  **MVC Architecture Support:** With built-in support for the Model-View-Controller (MVC) architecture, Laravel 8 provides a structured approach to application development.
6.  **Auto-complete Libraries and Configuration:** The framework offers comprehensive auto-complete libraries and configuration options, streamlining development and reducing errors.

| LMS | Company Name | Made Year |
|---|---|---|
| Moodle | Moodle Inc. | NA |
| Blackboard | Blackboard Inc. | 1997 |
| Canvas | Instructure lnc | 2008 |
| Sakai | Apereo Foundation | N/A |
| Latitude | Latitude Learning LLC | N/A |

**Table 1. Overall support platforms.**

### B.  Forum and Discussions
FLARUM was chosen based on several key factors:

Extensive Web Server Support: FLARUM is compatible with various web servers, including Apache (with mod_rewrite), Nginx, and Lighttpd.

PHP Compatibility: FLARUM supports PHP 5.5+ and requires specific extensions, including mbstring, pdo_mysql, openssl, json, gd, dom, and fileinfo.

Database Support: It supports MySQL 5.5+.

SSH Access: The provision of SSH (command-line) access is essential for managing and deploying the forum efficiently.

### C.  Analysis and specification of requirements
The functional requirements encompass the necessity for the research team to develop a new structure by studying existing OER systems and comparing technologies to build a new system with functionalities tailored for educators. These functions must be implemented by the development team. The proposed list of functional requirements includes:

1.  **Advanced Search:** An advanced search engine that allows users to search by author, keywords, description, category, material type, and other relevant criteria.
2.  **Semantic Search Engine:** This module enables users to conduct high-quality searches using keywords derived from document files such as PDFs, PowerPoints, descriptions, titles, and authors.
3.  **Browse Material:** A comprehensive "Browse All Materials" page that lists every available material. This feature enables users to focus their search on specific categories, material types, mobile filters, or other filters to refine the material list.
4.  **Registration:** A user registration system that allows individuals to become members, enabling them to create and upload content, participate in discussions, and access additional resources.

*   **Functional Requirements**
1.  **Bookmarks Collection:** Users can bookmark any material into a collection.
2.  **Forum:** A public medium or place used for debates, discussions, and questions.
3.  **Modify Content:** This module allows users to update or delete their uploaded content.
4.  **Contribute Material:** Members can contribute learning materials to the collection, defining metadata to help others find it.
5.  **Use Content Builder:** The Content Builder is an integrated web page and website development tool within the system, accessible to registered and logged-in members. It enables the creation of various designs, such as e-portfolios, lesson plans, pedagogical analyses, student reflections, online courses, tutorials, presentations, and community websites.
6.  **Ban Users:** Elite or admin users can ban others who spam the OER system.
7.  **Peer Review:** A module for rating and reviewing material, particularly by instructors in higher education institutions, indicating the quality and rating of their materials.

*   **Non-Functional Requirements:**
1.  **Ergonomics:** The front-end should be uniform, simple, and user-friendly, tailored to the ease of understanding and use, as educators are the primary audience for this system.
2.  **Performance:** The system should perform quickly and efficiently, processing tasks without errors, ensuring a smooth user experience.

| Tools | Moodle | Blackboard | Canvas | Latitude | Sakai |
|---|---|---|---|---|---|
| **Course Development Features** | | | | | |
| **Custom User Interface** | * | * | | * | * |
| **Custom Functionality** | * | * | | * | * |
| **Templates** | | * | | | |
| **Administrative  Features** | | | | | |
| **Administrative Reporting** | * | * | * | * | * |
| **Course Catalog** | * | * | * | * | * |
| **Grading** | * | * | * | * | * |
| **Defined User Roles** | * | | * | * | * |
| **Registration Management** | * | * | | * | * |
| **Mobile Access** | * | | * | | |
| **Data Import/Export** | * | * | * | * | |
| **Individual Plans** | * | * | | * | |
| **Collaboration Features** | | | | | |
| **Internal Messaging, Live Chat, Blog** | * | * | * | * | * |
| **File Exchange** | * | * | | | |
| **Discussion Forum** | * | * | * | * | * |
| **Collaboration Management** | * | | | | |
| **Assessment  Methods** | | | | | |
| **Skill Tracking** | * | * | | * | * |
| **Career Tracking** | * | * | * | * | * |
| **Instruction Methods** | | | | | |
| **E-Learning** | * | | | * | * |
| **Multimedia** | * | * | * | | * |
| **Virtual Classroom** | * | * | | * | * |

**Table 2.Comparison of the top LMS functionalities**

| | Bootstrap | Foundation |
|---|---|---|
| **Current version** | V5 | V6.0 |
| **Details** | Bootstrap is recognized as the most preferred HTML, CSS, and JavaScript framework for crafting responsive, mobile-first projects across the web. It stands as the foremost advanced responsive front-end framework worldwide. | It stands as the foremost advanced responsive front-end framework worldwide. |
| **Preprocessors** | Less and Sass | Sass |
| **Responsive, Modular, Templates** | Yes | Yes |
| **Unique components** | Jumbotron | on Bar, Clearing Lightbox, Flex Video, Keystrokes, Joyride, and Pricing Tables. |
| **Icon Set** | Glyphicons Halflings set | Foundation Icon Fonts |
| **Extras/Add-ons** | While Bootstrap does not come bundled with extras, it offers compatibility with various third-party plugins. On the other hand, Foundation Icon Fonts does include certain extras and add-ons. | Yes |
| **Customization** | Bootstrap provides a basic GUI Customizer, requiring manual input for color values. | Customization is done manually alone; no GUI is used. |
| Browser support | IE8+ (you require Respond.js for IE8), Firefox, Chrome, and Safari | Windows Phone 7+, iOS, Android, and Firefox, Chrome, Safari, and IE9+ |

**Table 3. Front End Framework**

## 5.  CONCLUSIONS
The development of an Open Education Resources (OER) system for educational institutions requires a comprehensive understanding of both functional and non-functional requirements. Our evaluation of functional necessities has highlighted the need for advanced search capabilities, semantic search engines, material browsing, user registration, member review, bookmarking, forum functionality, content modification and contribution, content building, user banning, and peer review. These functionalities are designed to cater to the diverse needs of educators and learners, aiming to create a robust and user-friendly platform for the dissemination and utilization of educational resources.

Moreover, the non-functional requirements emphasize the importance of an ergonomic and user-friendly interface, high system performance, and efficient maintenance. These considerations are vital in guaranteeing that the OER system is not only technologically robust but also user-centered and easily maintainable, aligning with the requirements and expectations of educators and learners.

## 6.  DISCUSSION
The development of an OER system is a complex and multifaceted undertaking, requiring a delicate balance between functional and non-functional requirements. The functionality of the system needs to cater to the diverse needs of educators and learners, providing advanced search capabilities, user-friendly content creation and modification tools, and an interactive platform for discussion and collaboration. The inclusion of semantic search engines and user review functionalities further enhances the platform's usability and effectiveness.

In parallel, the non-functional requirements play a pivotal role in ensuring that the system is not only efficient and high-performing but also user-friendly and easy to maintain. The ergonomics of the system are particularly vital, as educators are the primary target audience. The system's performance and ease of maintenance are equally important, ensuring that it can effectively support the evolving needs of educational institutions and their users.

Ultimately, the successful development of an OER system for educational institutions necessitates a holistic approach, considering both the functional and non-functional aspects to create a comprehensive, user-centric, and high-performing platform. By addressing these requirements, the OER system can effectively support the educational needs of institutions and contribute to a more accessible and interactive learning environment for educators and learners alike.